\documentclass[pdflatex,sn-mathphys-num]{sn-jnl}

\usepackage{graphicx}%
\usepackage{multirow}%
\usepackage{amsmath,amssymb,amsfonts}%
\usepackage{amsthm}%
\usepackage{mathrsfs}%
\usepackage[title]{appendix}%
\usepackage{xcolor}%
\usepackage{textcomp}%
\usepackage{manyfoot}%
\usepackage{booktabs}%
\usepackage{algorithm}%
\usepackage{algorithmicx}%
\usepackage{algpseudocode}%
\usepackage{listings}%
\usepackage{subcaption}

\theoremstyle{thmstyleone}%
\theoremstyle{thmstyletwo}%

\theoremstyle{thmstylethree}%

\definecolor{color1}{RGB}{0,0,120}

\begin{document}

\title{Brazil Data Commons: A Platform for Unifying and Integrating Brazil's Public Data}


\author*[1]{\fnm{Isadora} \sur{Cristina}}\email{isadorarodrigues@dcc.ufmg.br}

\author[1,2]{\fnm{Ramon} \sur{G. Gonze}}\email{ramon.gonze@inria.fr}

\author[1]{\fnm{Jônatas} \sur{Santos}}\email{jonatash118@gmail.com}

\author*[3]{\fnm{Julio} \sur{C. S. Reis}}\email{jreis@ufv.br}

\author[1]{\fnm{Mário} \sur{S. Alvim}}\email{msalvim@dcc.ufmg.br}

\author[4]{\fnm{Bernardo} \sur{L. Queiroz}}\email{blanza@ufmg.br}

\author[1]{\fnm{Fabrício} \sur{Benevenuto}}\email{fabricio@dcc.ufmg.br}

\affil*[1]{\orgdiv{Departament of Computer Science}, \orgname{Universidade Federal de Minas Gerais}, \orgaddress{\street{Av. Pres. Antônio Carlos}, \city{Belo Horizonte}, \postcode{31270-901}, \state{Minas Gerais}, \country{Brazil}}}

\affil[2]{\orgdiv{Department of Computer Science}, \orgname{École Polytechnique de Paris}, \orgaddress{\street{Rte de Saclay}, \city{Palaiseau}, \postcode{91120}, \state{Île-de-France}, \country{France}}}

\affil[3]{\orgdiv{Department of Informatics}, \orgname{Universidade Federal de Viçosa}, \orgaddress{\street{Av. P. H. Rolfs}, \city{Viçosa}, \postcode{36570-900}, \state{Minas Gerais}, \country{Brazil}}}

\affil[4]{\orgdiv{Department of Demography}, \orgname{Universidade Federal de Minas Gerais}, \orgaddress{\street{Av. Pres. Antônio Carlos}, \city{Belo Horizonte}, \postcode{31270-901}, \state{Minas Gerais}, \country{Brazil}}}


\abstract{The fragmentation of public data in Brazil, coupled with inconsistent standards and limited interoperability, hinders effective research, evidence-based policymaking and access to data-driven insights. To address these issues, we introduce Brazil Data Commons, a platform that unifies various Brazilian datasets under a common semantic framework, enabling the seamless discovery, integration and visualization of information from different domains. By adopting globally recognized ontologies and interoperable data standards, Brazil Data Commons aligns with the principles of the broader Data Commons ecosystem and places Brazilian data in a global context. Through user-friendly interfaces, straightforward query mechanisms and flexible data access options, the platform democratizes data use and enables researchers, policy makers, and the public to gain meaningful insights and make informed decisions. This paper illustrates how Brazil Data Commons transforms scattered datasets into an integrated and easily navigable resource that allows a deeper understanding of Brazil’s complex social, economic and environmental landscape.}





\maketitle

\section{Introduction}\label{sec1}

Tackling pressing social challenges—such as environmental crises, large-scale pandemics and complex economic issues—requires sound research to understand reality and inform policy actions \cite{passos2022, saliba2023}. Such research, in turn, depends on high-quality and reliable data \cite{herrera2007improving}. Fortunately, a substantial amount of important data is already available, as democratic nations have witnessed a significant increase in open data dissemination over the past few decades. This trend, driven by the expansion of information and communication technologies \cite{bertot2010}, has encouraged countries to commit to initiatives such as the Open Government Partnership \cite{correa2014}, which aim to foster transparency and facilitate the distribution of data in electronic formats to a broad range of users.

However, although open data is, by definition, free to use without restrictions \cite{correa2014}, this does not necessarily mean that governmental open data is easily accessible or usable \cite{dang2023statistical}. Sophisticated research endeavors aimed at addressing complex questions often require data that cut across multiple domains and are provided by different government agencies. Still, the integration of distinct and large-scale datasets remains a challenge that has yet to be fully addressed. Many existing initiatives fall short of delivering a seamless and user-friendly experience, particularly when it comes to navigating and linking heterogeneous sources. As a result, researchers often spend a significant portion of their time on repetitive tasks such as data cleaning, formatting, and integration \cite{press2016} —efforts further complicated by the lack of standardized data formats \cite{deoliveira2018}. For non-specialists lacking technical skills, these challenges become even more formidable \cite{martins2013}.

In Brazil, a country with continental dimensions and extensive public datasets, open data initiatives face significant challenges \cite{shikida2021guia, passos2022, silva2025open, oliveira2024open}. As one of the founding members of the Open Government Partnership, Brazil has long been committed to transparency as a means of combating corruption and strengthening public trust \cite{breitman2012}. The \textit{Lei de Acesso à Informação} (LAI) \cite{BrazilLAI2011}, enacted in 2011, pressured government agencies to release data for public use. Nevertheless, each agency typically focuses on a single domain and retains considerable autonomy in deciding how data are published, with no enforced standardization. The health sector, for example, revealed during the COVID-19 pandemic the limitations of Brazil’s data infrastructure \cite{passos2022}: a study identified 54 national-level information systems, each fed by municipalities throughout the country \cite{coelho2021}. This fragmentation results in incomplete and inconsistent datasets, underscoring the need for a common vocabulary and greater interoperability \cite{deoliveira2018}.

Beyond the absence of standardization, privacy and data protection have also become major concerns in the management of public information in Brazil, with evidence emerging across multiple domains. In the health sector, during the COVID-19 pandemic, the federal government launched a platform originally intended for long-term implementation, but deployed hastily, while little transparency has been provided regarding the protection of data, especially concerning how private-sector access is regulated \cite{passos2022, rnds2025}. Furthermore, when compared with disclosure policies adopted by other nations—particularly in the field of education, for example—the \textit{Instituto Nacional de Estudos e Pesquisas Educacionais Anísio Teixeira} (INEP) \cite{inep2025} releases an unusually large amount of microdata, often with a high level of detail. The techniques used to mitigate potential privacy risks are also relatively weak when compared with international initiatives \cite{alvim2020}.

Outside Brazil, there are several international efforts to map and make available data about the country, such as those led by platforms like Eurostat \cite{eurostat2025} and the World Bank \cite{worldbank2025}. However, these initiatives often fall short of what is needed for in-depth analysis. While they are useful for global comparisons, their limited granularity constrains more detailed regional and sectoral assessments of Brazil — a limitation that is particularly problematic in a country of continental scale, where socioeconomic, demographic, and environmental conditions vary widely across regions.

In practical terms, the challenges discussed above have tangible implications for those who rely on public data in Brazil. Consider, for instance, a health economist investigating the socioeconomic determinants of infant mortality across Brazilian municipalities. Such an analysis would require combining mortality indicators from DataSUS \cite{datasus2025}, demographic and income data from the \textit{Instituto Brasileiro de Geografia e Estatística} (IBGE) \cite{ibge2025}, and historical series on public health investment from IpeaData \cite{ipea2019}. Traditionally, this process involves identifying the relevant sources, handling disparate formats, reconciling inconsistent municipal codes and territorial aggregations that have changed over time, and developing custom scripts for each stage—a workflow that can take weeks or even months before the actual analysis begins. 

To address these limitations, we present Brazil Data Commons, available at \url{https://brazildatacommons.com.br}. Brazil Data Commons consolidates a wide range of Brazilian public datasets—spanning socioeconomic indicators, environmental statistics, and more—into a unified system designed to promote interoperability. Its most distinctive feature lies in the adoption of a clear and globally recognized ontology, ensuring semantic consistency across datasets. In a country as large and institutionally diverse as Brazil, where municipalities operate with considerable digital autonomy, adherence to a shared semantic framework is essential. By aligning with international standards, Brazil Data Commons lays the groundwork for harmonized data practices nationwide, regardless of local implementation variations. Importantly, the platform's data pipeline includes systematic privacy risk assessments for microdata, ensuring that it does not compromise individual privacy protections. Furthermore, to ensure accessibility, Brazil Data Commons also provides both graphical interfaces and standardized APIs, enabling users—from developers to the general public—to interact with complete and structured data using minimal programming skills.

While firmly grounded in the Brazilian context, Brazil Data Commons draws upon the foundational principles of the global Data Commons ecosystem \cite{guha2023}. It leverages globally adopted ontologies—endorsed by indexing platforms such as Google \cite{google2025}, Microsoft \cite{microsoft2025}, Pinterest \cite{pinterest2025}, and Yandex \cite{yandex2025}—and implemented by over 45 million websites worldwide \cite{schemaorg2015}. This semantic alignment transforms fragmented and siloed datasets into a cohesive and interoperable resource, facilitating streamlined data management and analysis. Moreover, Brazil Data Commons’s integration into the broader Data Commons network fosters a collaborative API ecosystem and allows Brazilian datasets to be contextualized within international research frameworks—helping position the country within the landscape of globally relevant, data-driven studies \cite{blicharska2017}.

To demonstrate the relevance and practical applicability of Brazil Data Commons, we explore four distinct use cases that illustrate how data from different Brazilian databases can be integrated and analyzed to provide actionable insights for researchers, practitioners, and decision-makers. These cases range from basic descriptive analyses—such as examining trends in monthly labor income across demographic groups and regions—to more sophisticated applications, including fine-grained spatial analysis of life expectancy across more than 5,000 Brazilian municipalities, intuitive visualizations comparing education indicators over time, and international benchmarking of socioeconomic variables such as years of schooling. Critically, all these analyses can be conducted directly through the platform's interface, as the labor-intensive work of collecting, cleaning, and harmonizing data from disparate sources with heterogeneous vocabularies has already been performed. This infrastructure enables evidence-based research at multiple scales, from local interventions to cross-national comparative studies, without requiring users to possess advanced technical skills or invest significant time in data preparation.

By offering user-friendly tools, simplified search mechanisms, and ready-made visualizations, Brazil Data Commons further democratizes access to high-quality public data. It empowers researchers, policymakers, journalists, and civil society to derive insights and make informed decisions without requiring advanced technical expertise. In this article, we explore how Brazil Data Commons contributes to overcoming longstanding barriers in Brazil’s data ecosystem—transforming dispersed information into a coherent and actionable resource essential for evidence-based governance and social progress.

The remainder of this article is structured as follows: we first situate Brazil Data Commons in relation to similar initiatives, discussing how it differs from existing solutions and the gaps it seeks to address. We then present the system itself, describing the architecture of the web platform and the tools used for data acquisition. Next, we illustrate how users can interact with the platform through concrete use cases. This is followed by a section on ethical and privacy considerations incorporated into the development of the tool. Finally, we reflect on the results of our work, its limitations, and potential future directions.

\section{Related Work}\label{sec2}

Next, we highlight existing initiatives that aim to facilitate the discovery, access, and reuse of public data. We examine both international and Brazilian efforts, with particular attention to projects that include Brazilian data at different territorial levels. By analyzing their approaches, scopes, and limitations, we discuss how our initiative differs from others and addresses key gaps in semantic integration and distributed data access.

\subsection{Governmental Open Data Portals}

The challenge of democratizing public data access in Brazil has been addressed through a variety of strategies. The Brazilian Open Data Portal (\url{dados.gov.br}) \cite{dadosgovbr2025} emerged within the global wave of national open data portals—following pioneering efforts such as \url{data.gov} \cite{datagov2025} in the United States and \url{data.gov.uk} \cite{datagovuk2025} in the United Kingdom. This government-led initiative serves primarily as a metadata catalog. While it plays an important role in enabling dataset discovery, it does not support data integration or semantic interoperability \cite{lourenco2015}. 

Beyond this portal, several domain-specific initiatives illustrate Brazil’s tradition of making public data available. Agencies such as IBGE \cite{ibge2025}, Brazilian Ministry of Health (Datasus) \cite{datasus2025} and INEP \cite{inep2025} provide their own APIs and data portals, while the \textit{Instituto de Pesquisa Econômica Aplicada} (IPEA) \cite{ipea2019} maintains IPEAData, a long-standing macroeconomic, regional, and social database. IPEAData offers detailed documentation and methodological guidelines, enabling longitudinal analyses of specific topics over extended periods. However, combining information across domains still requires additional programming and data cleaning. Taken together, these platforms deliver valuable data within their respective scopes, but none adopt shared data models or facilitate cross-domain linkage, thereby constraining integrated analysis.

\subsection{Research-Oriented Platforms and Datasets}

Other Brazilian initiatives attempt to address these limitations through alternative strategies. Dataset repositories such as DataViva \cite{freitas2023} and Base dos Dados are both examples of that. DataVida offers a curated interface for accessing and visualizing economic data. Its strength lies in user-friendly dashboards and graphical visualizations, making complex datasets more accessible to the public. However, it remains narrowly focused on economic indicators and does not support the integration of datasets from other domains such as education or health. On the other hand, Base dos Dados is an initiative focused on centralizing public datasets into a unified PostgreSQL and BigQuery-based infrastructure \cite{basedosdados2025}. It provides queryable datasets in formats that align with data science workflows, with a particular emphasis on microdata. While this focus greatly expands analytical possibilities, it also requires users to perform additional steps in data preparation. Although both platforms represent clear improvements in usability and access compared to many official sources, they do not incorporate formal ontologies, entity resolution mechanisms, or a semantic layer. 

Focusing more narrowly on municipal-level information, BrStats \cite{toledo2023brstats} develops a database that compiles administrative records from different Brazilian agencies. Unlike broader infrastructures such as DataViva or Base dos Dados, BrStats does not operate as a full platform but rather as a structured dataset. It organizes indicators related to population, economic conditions, and child health at the city level, thereby expanding the availability of disaggregated statistics for local analyses. While this emphasis on microdata and municipal granularity enhances analytical potential, it also introduces challenges in terms of standardization, comparability, and integration with broader national datasets. 

A different approach is taken by the Plataforma de Dados Desidentificados, which  provides access to large-scale, anonymized, and linked datasets on social determinants of health, including the 100 Million Brazilian Cohort \cite{pdd2025}. It constitutes a significant research infrastructure for public health studies, leveraging privacy-preserving record linkage techniques. In this sense, it represents a rare example of a platform enabling linked data practices in Brazil. However, access is typically restricted to researchers under strict protocols, and the data remains narrowly focused on the health domain.

\subsection{International Data Platforms}

International initiatives such as the World Bank Open Data \cite{worldbank2025}, OECD Data Portal \cite{oecd2025}, and Eurostat \cite{eurostat2025} provide standardized datasets that often include national-level indicators for Brazil. These platforms typically employ consistent schemas and support comparisons across countries. Nevertheless, their coverage of Brazilian data tends to be coarse-grained—aggregated at the national level—and lacks the granularity needed for regional or municipal analysis. Moreover, they rarely integrate with local data infrastructures, limiting their utility for nuanced, territory-specific insights.

\subsection{Research Gap}

Unlike the approaches discussed above, Brazil Data Commons not only eliminates the need for users to manually collect, clean, and merge public datasets, but also makes these datasets readily available for download, exploration, and reuse. Through a user-friendly interface and graphical visualizations, the platform enables intuitive access to structured data while promoting consistent standards for data representation. Crucially, it overcomes common limitations found in other initiatives—such as the absence of shared schemas, lack of semantic annotation, and the unavailability of open semantic APIs—which often hinder interoperability and prevent integrated analysis across domains. By aligning with the broader Data Commons ecosystem and adopting a shared semantic infrastructure, Brazil Data Commons facilitates knowledge representation, higher-level reasoning, and integration across both thematic and territorial dimensions—addressing a key gap in the Brazilian open data landscape.

\section{Architecture}\label{sec3}

The Brazil Data Commons architecture is designed to unify diverse datasets under a shared semantic framework. To achieve this, it combines a distributed web platform with a semantic ETL pipeline. The web platform provides the interaction layer, enabling users and external systems to query, visualize, and consume data through standardized APIs and modular components. Complementing it, the ETL pipeline ingests, normalizes, and transforms raw information—sourced from government portals and institutional repositories—into semantically consistent structures that can be seamlessly integrated into the broader data framework, a knowledge graph. Together, these two dimensions establish a self-sustaining ecosystem: while the platform exposes data in an accessible and interoperable way, the pipeline guarantees the quality, coherence, and continuous update of the underlying repository.

\subsection{Web Platform}

Instead of functioning as a large, centralized repository, Data Commons is manifested as a distributed, interoperable network of APIs \cite{guha2023}. At the heart of its architecture is a set of well-defined schemas—a superset of Schema.org \cite{guha2016}, itself an extension of the Resource Description Framework (RDF) \cite{brickley1998}—that describe entities, properties, and relationships among data. These schemas serve as a ``shared vocabulary'', ensuring that data from different sources can be coherently combined and referenced by any API in the network, even if they were built in completely different ways. This integration results in a knowledge graph: a set of nodes and edges representing entities (such as countries, states, cities) and their properties (population, area, Human Development Index \cite{undp2025}). It is worth noting that Data Commons relies on the principle of Reference by Description \cite{guha2015}, meaning that instead of arbitrary identifiers, clear and standardized semantic descriptions are used. For example, rather than a simple ``ID 1234'' for a place, the reference is made through a description that clarifies which entity is being referred to (e.g., ``The city of Belo Horizonte in Brazil''). This semantic approach facilitates human understanding and does not require different data sources to share the same identifiers.

\begin{figure*}[h]
  \centering
  \includegraphics[width=1.05\linewidth]{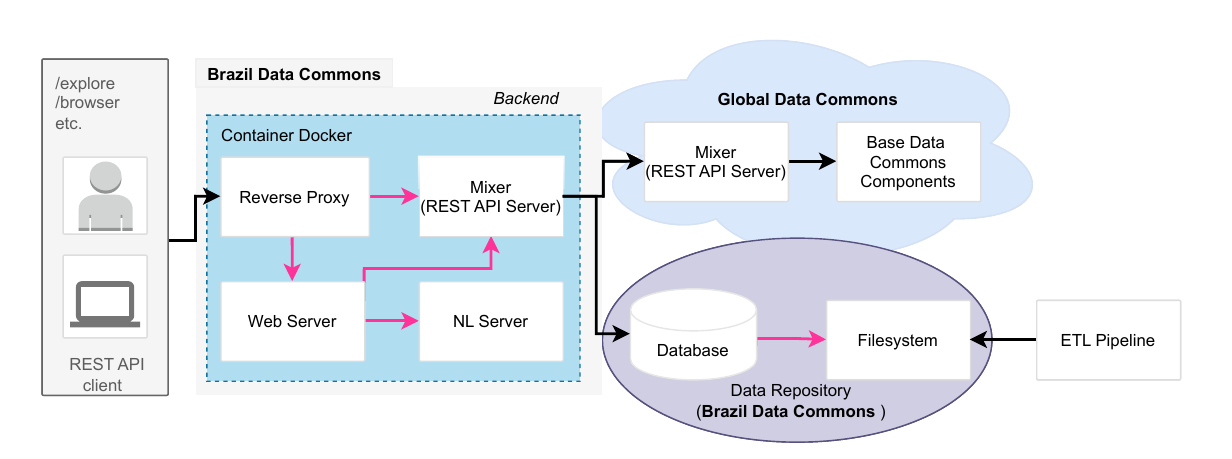}
  \caption{Brazil Data Commons architecture.}
  \label{fig:architecture}
\end{figure*}

Brazil Data Commons has a modular architecture that integrates customized components with the network’s base infrastructure. The system consists of key elements as shown in Figure \ref{fig:architecture}: the reverse proxy, which routes external requests to internal servers; the Web Server, responsible for the user interface; the NL Server, which enables natural language queries; and the integration component (``Mixer''), which extends beyond data integration, operating as a REST API server that bridges the data repository with end users. In the following paragraphs, each of these components is described in greater detail.

The Web Server is responsible for rendering the graphical interface and issuing the necessary API calls to generate visualizations, including the construction of complex analytical views and support tools such as the ``Data Download Tool.'' It leverages the NL Server to enhance the user experience by translating natural language requests into executable queries. Operating behind the reverse proxy, it also benefits from request routing and load balancing mechanisms, which ensure that traffic from the browser is efficiently distributed and the system remains resilient under heavy demand. 

The Mixer is the system’s core component. It handles front-end requests using locally stored data, ensuring they are properly formatted, validated, and integrated. Beyond this, it enriches responses by dynamically combining local information with external datasets when required—hence the origin of its name. In addition to serving data to the Web Server, the Mixer also constitutes a public-facing REST API, enabling external tools and users to access the integrated knowledge graph in a standardized and interoperable way. By centralizing integration logic and exposing a unified interface, the Mixer ensures consistency, scalability, and reusability across the entire ecosystem.

Unlike the standard Data Commons implementation, which relies heavily on cloud services, the entire Brazil Data Commons stack—including both the web platform and the data repository—has been specifically adapted to operate in local computing environments. This local-first design enables institutions to deploy and maintain their own instances of the system without cloud dependencies, reducing the need for resources and thereby broadening potential adoption across diverse contexts. 

\subsection{Semantic ETL: Data Ingestion and Transformation}

Fed by the semantic ETL pipeline, the data repository constitutes the core storage layer of Brazil Data Commons. It is composed of a file system, where raw source files, obtained from government portals and institutional databases, are maintained, and a database, which is incrementally built from these files. During ingestion, the datasets are normalized and transformed to adhere strictly to the standardization proposed by Data Commons, ensuring full compatibility with the analytical tools—such as dynamically generated visualizations—available across the network. The database’s incremental design reduces computational and operational costs for each update cycle, thereby enhancing the pipeline’s usability and encouraging contributions from multiple sources.

\begin{figure*}[h]
  \centering
  \includegraphics[width=1.05\linewidth]{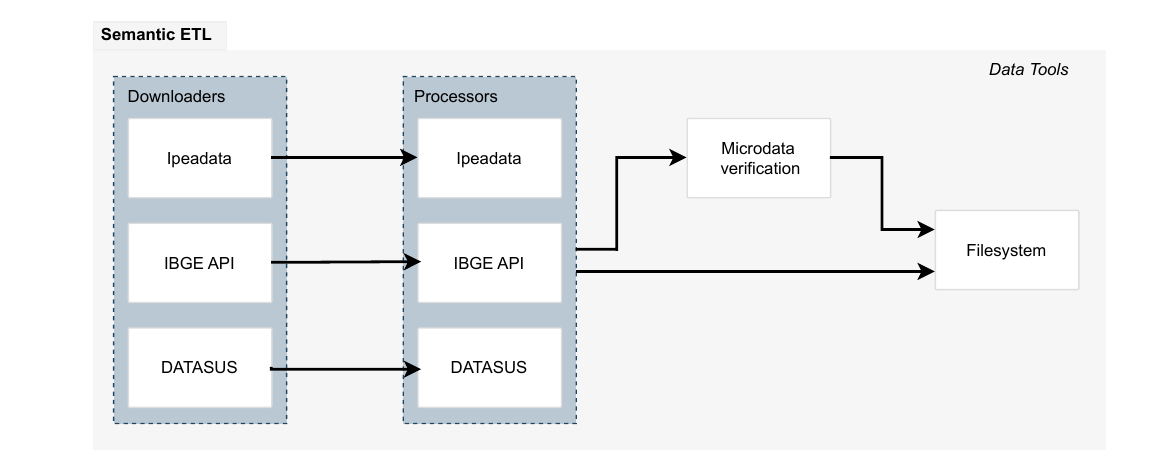}
  \caption{Brazil Data Commons Semantic ETL.}
  \label{fig:etl}
\end{figure*}

Another important aspect of the pipeline is its commitment to privacy and data protection. Particular attention is given to datasets containing microdata, that is, detailed information at the individual level, as these pose significant privacy risks. During data ingestion, the pipeline performs automated checks to identify potential vulnerabilities. Specifically, it assesses the risks of re-identification, when a record can be successfully linked to its owner, and attribute inference, when the value of a sensitive attribute of an individual can be predicted or revealed.
Evidence from Brazilian health datasets indicates that approximately 30\% of the population faces a 90\% risk of having their sensitive attributes inferred, as shown by Gonze et al. (2025). The definition of an attribute inference attack used in the automated check is formally provided in \cite{gonze2025riscos}. Re-identification operates in a similar manner to attribute inference; however, instead of focusing on the leakage of a sensitive attribute, it concerns whether a record can be linked back to its owner. Datasets with such high exposure risks are excluded from publication in the current version of the repository.

It is important to note that the definition of sensitive information can vary according to the specific assets or data intended for protection.
We have manually classified sensitive information based on the \textit{Lei Geral de Proteção de Dados Pessoais} (LGPD)~\cite{BrazilLGPD2018}, that states in its $5^{th}$ article: ``\textit{sensitive personal data: personal data about racial or ethnic origin, religious belief, political opinion, membership of a trade union or organization of a religious, philosophical or political nature, data relating to health or sexual life, genetic or biometric data, when linked to a natural person}''.
Although these data are not yet included, future work will focus on developing and integrating mitigation techniques, such as anonymization and differential privacy, to enable their safe use while preserving analytical value.

Beyond serving as the backbone of the repository, the ETL modules represent a contribution in their own right. By automating the extraction and normalization of datasets from heterogeneous government sources, they reduce the effort required for manual data curation and can be reused independently by researchers or institutions interested only in streamlined access to standardized data. This modularity extends the impact of Brazil Data Commons, ensuring that even partial adoption of its components can foster data accessibility and reuse.

\subsection{Data Repositories}

This section lists the portals from which the platform’s data was collected and integrated.

\vspace{2mm}

\textbf{Ipeadata.} As discussed earlier, Ipeadata \cite{ipea2019} is a longstanding initiative maintained by the Institute for Applied Economic Research (IPEA), a federal government institution, that aggregates statistical and historical series covering economic, demographic, geographic, and social indicators at multiple territorial levels in Brazil. The database includes records dating back to 1872, making it one of the most comprehensive collections of time-series data in the country. More than three thousand series are currently accessible through a public REST API; while the source exposes additional territorial breakdowns, Brazil Data Commons standardizes and maintains data only at the country, state (UF), and municipality levels. Much of this data originates from official institutions such as IBGE \cite{ibge2025}, the Central Bank, and the National Treasury, positioning Ipeadata as a central aggregator for socioeconomic indicators in Brazil.

\vspace{2mm}

\textbf{IBGE API.} Maintained by the Brazilian Institute of Geography and Statistics (IBGE) \cite{ibge2025}, this API provides access to aggregated data from official surveys and censuses. It exposes thousands of statistical tables covering demographic, social, and economic indicators, with flexible parameters for selecting variables, territorial levels, time periods, and classifications. Data can be queried at multiple granularities, from the national scale down to municipalities and, in some cases, even census tracts. 

\vspace{2mm}

\textbf{DATASUS.} Maintained by Brazil’s Ministry of Health, the OpenDataSUS portal aggregates a wide range of public health datasets, with 67 data collections spanning domains such as surveillance, vaccination, primary care, and health indicators \cite{datasus2025}. Many datasets are derived from official surveillance systems, such as SINAN (Notifiable Disease Information System) \cite{sinan2025}, SI-PNI (National Immunization Program) \cite{sipni2025}, and SISAGUA (Water Quality Monitoring) \cite{sisagua2025}, reflecting real-time and historical public health metrics. The dataset catalog provides metadata, documentation, and usage manuals (e.g., for COVID‑19 vaccination data), enabling standardized ingestion across tools and platforms.

\section{Results and Use Cases}\label{sec4}

To illustrate the practical value of Brazil Data Commons, this section is organized into two parts. First, we provide an overview of the platform’s interface and technical features that facilitate access to structured data, including standardized APIs and intuitive visualization tools. Next, we present selected use cases that highlight the analytical possibilities enabled by Brazil Data Commons—ranging from basic descriptive analyses to international comparisons and spatial assessments—demonstrating how the platform supports both specialized research and evidence-based public discourse.

\subsection{Interface}

\begin{figure}[h]
    \centering
    \includegraphics[width=\linewidth]{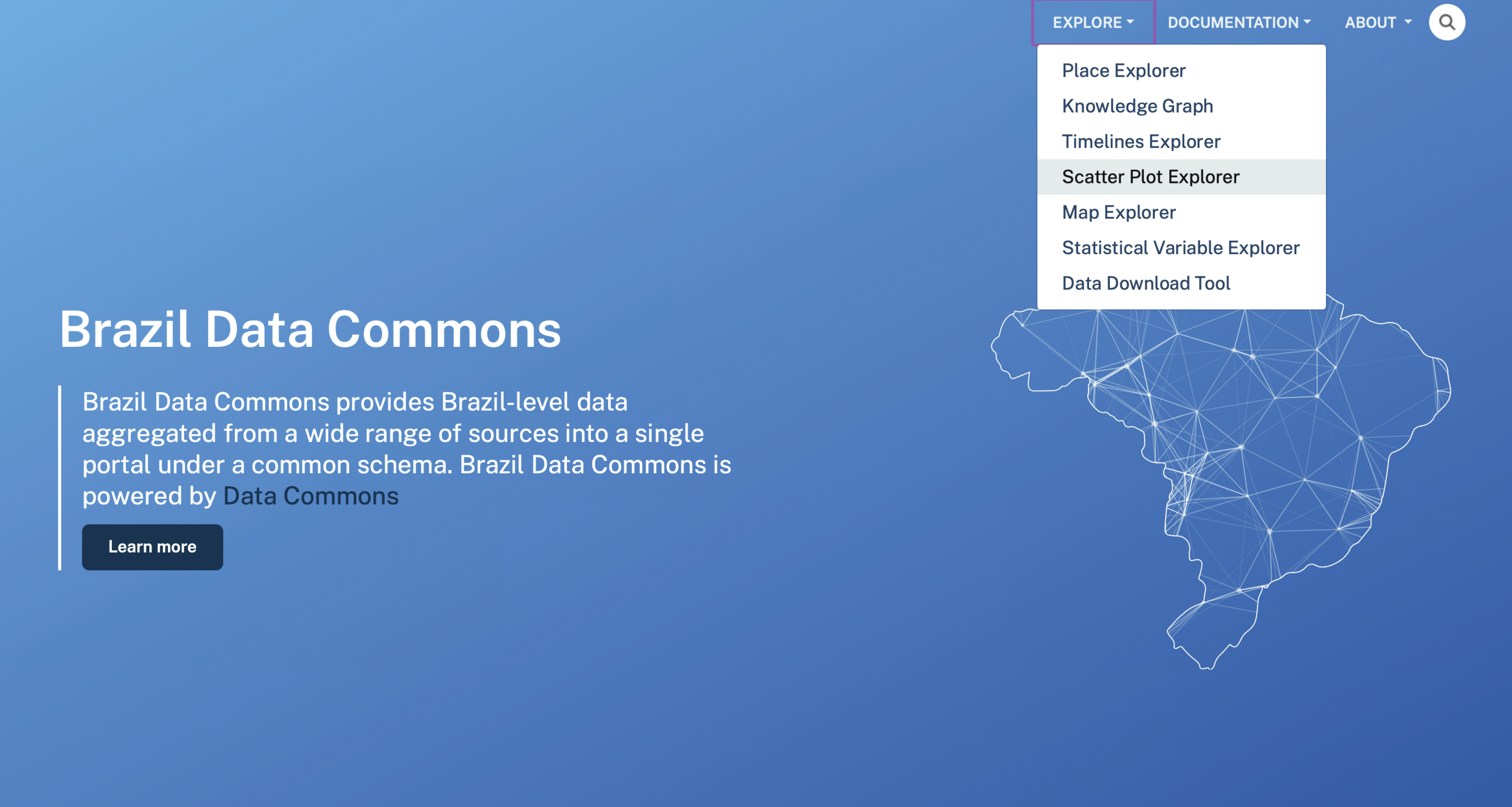}
    \caption{Screenshot of Brazil Data Commons: Homepage.}\label{fig:home}
\end{figure}

\begin{figure}[h]
    \centering
    \includegraphics[width=\linewidth]{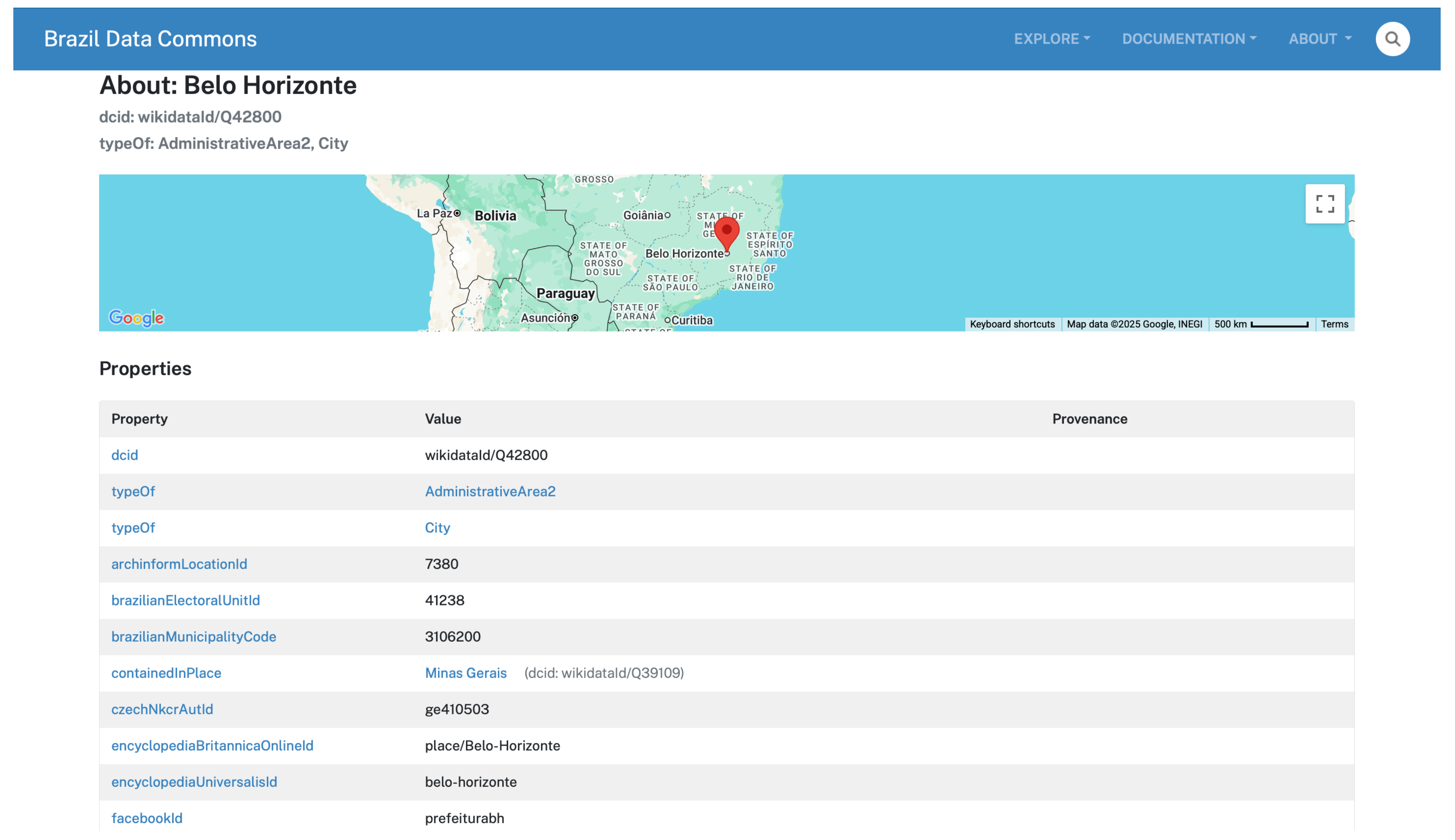}
    \caption{Screenshot of Brazil Data Commons: Knowledge Graph.}\label{fig:knowledgegraph}
\end{figure}

The set of figures in this section presents snapshots of the Web interface of the Brazil Data Commons. Figure \ref{fig:home} shows the homepage, where users can access the main tools through the ``Explore'' menu, such as the place explorer and graphical visualization modules. Figure \ref{fig:knowledgegraph} illustrates the representation of a node in the knowledge graph—in this example, the city of Belo Horizonte—through which users can retrieve detailed information about any node in the graph.

\begin{figure}[h]
    \centering
    \includegraphics[width=\linewidth]{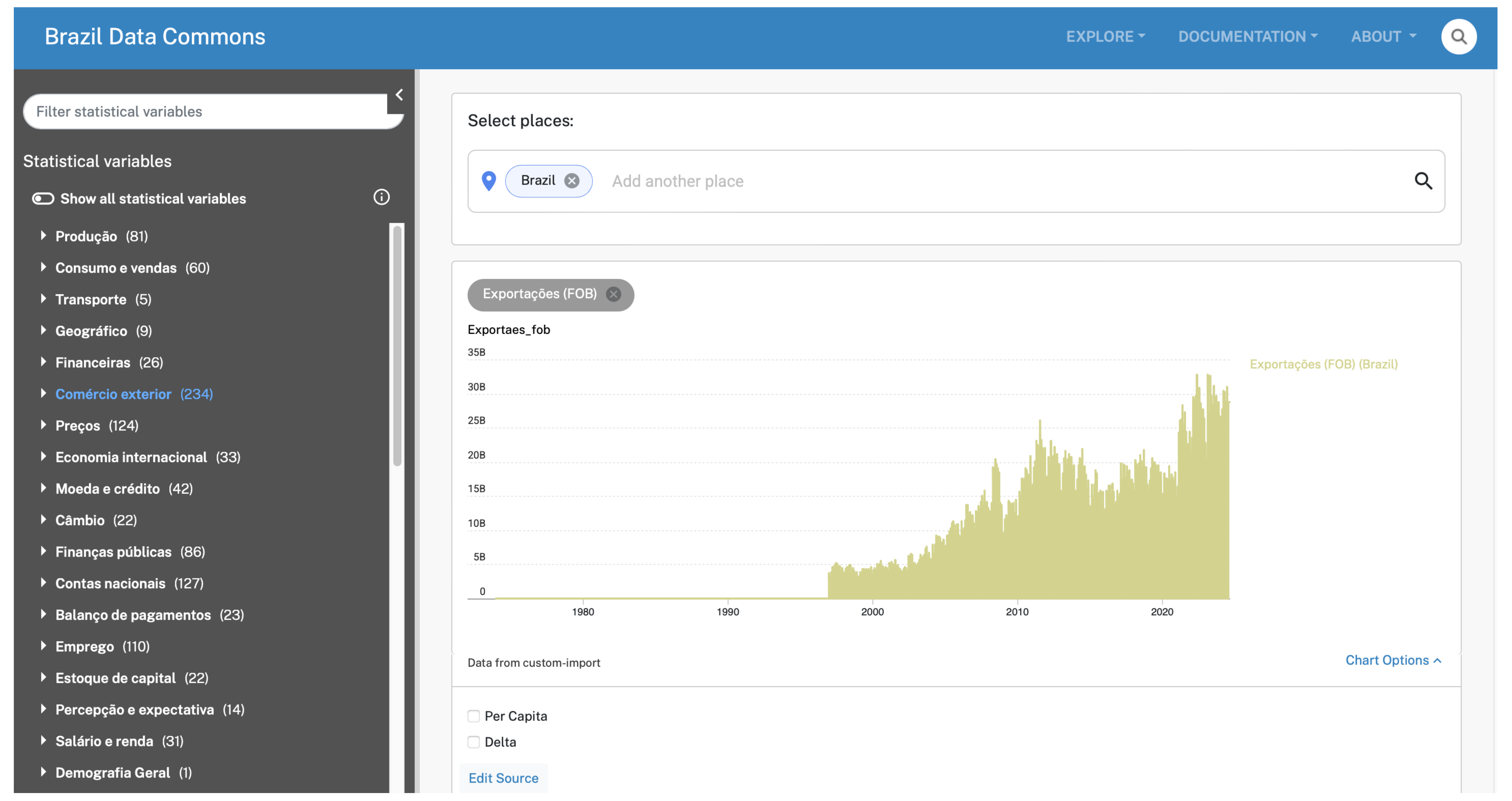}
    \caption{Screenshot of Brazil Data Commons: Timeline.}\label{fig:timeline}
\end{figure}

\begin{figure}[h]
    \centering
    \includegraphics[width=\linewidth]{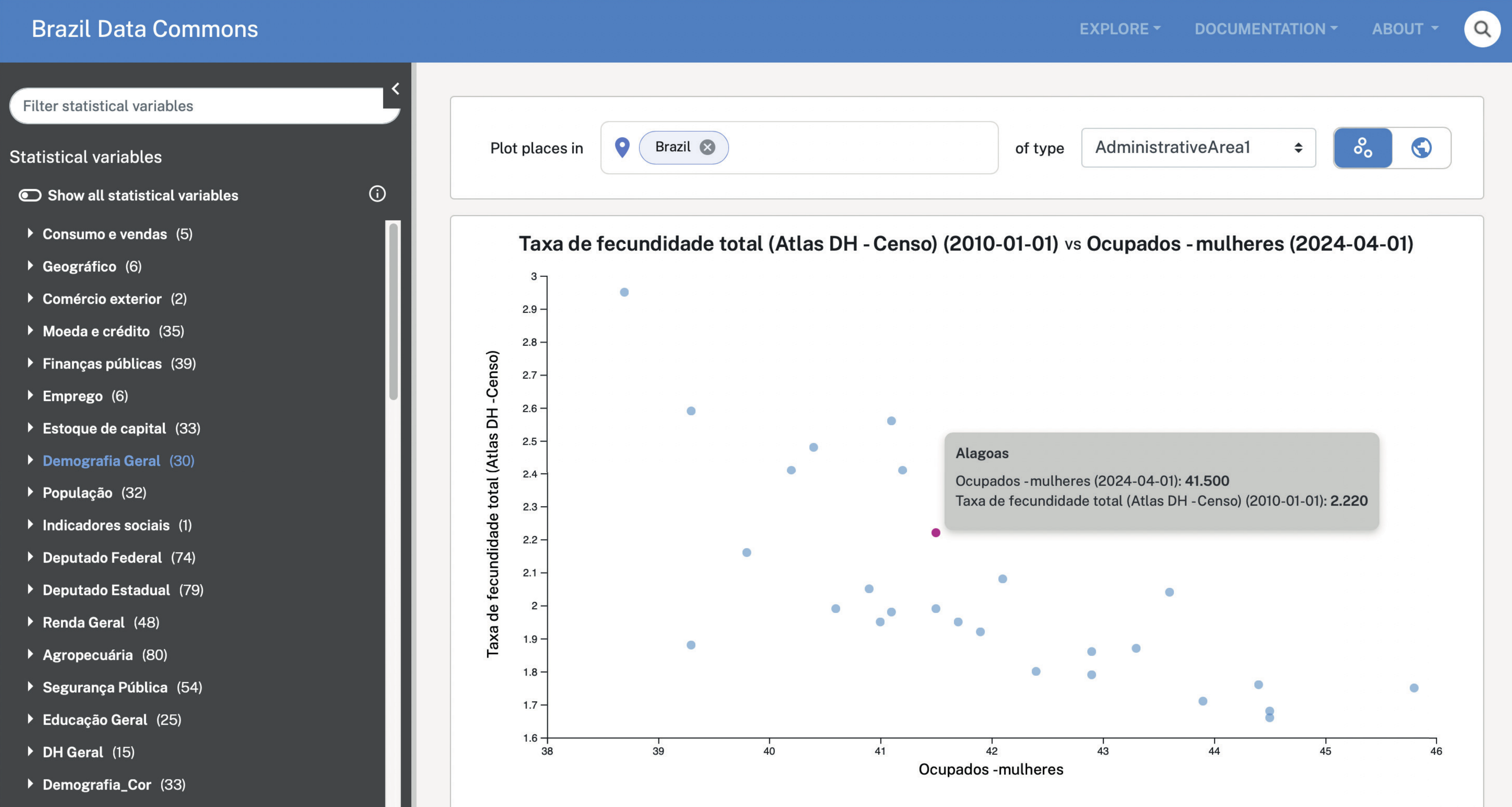}
    \caption{Screenshot of Brazil Data Commons: Scatter plot of `Total Fertility Rate' vs. `Employed Women' among all Brazilian states.}\label{fig:scatter}
\end{figure}

\begin{figure}[h]
    \centering
    \includegraphics[width=\linewidth]{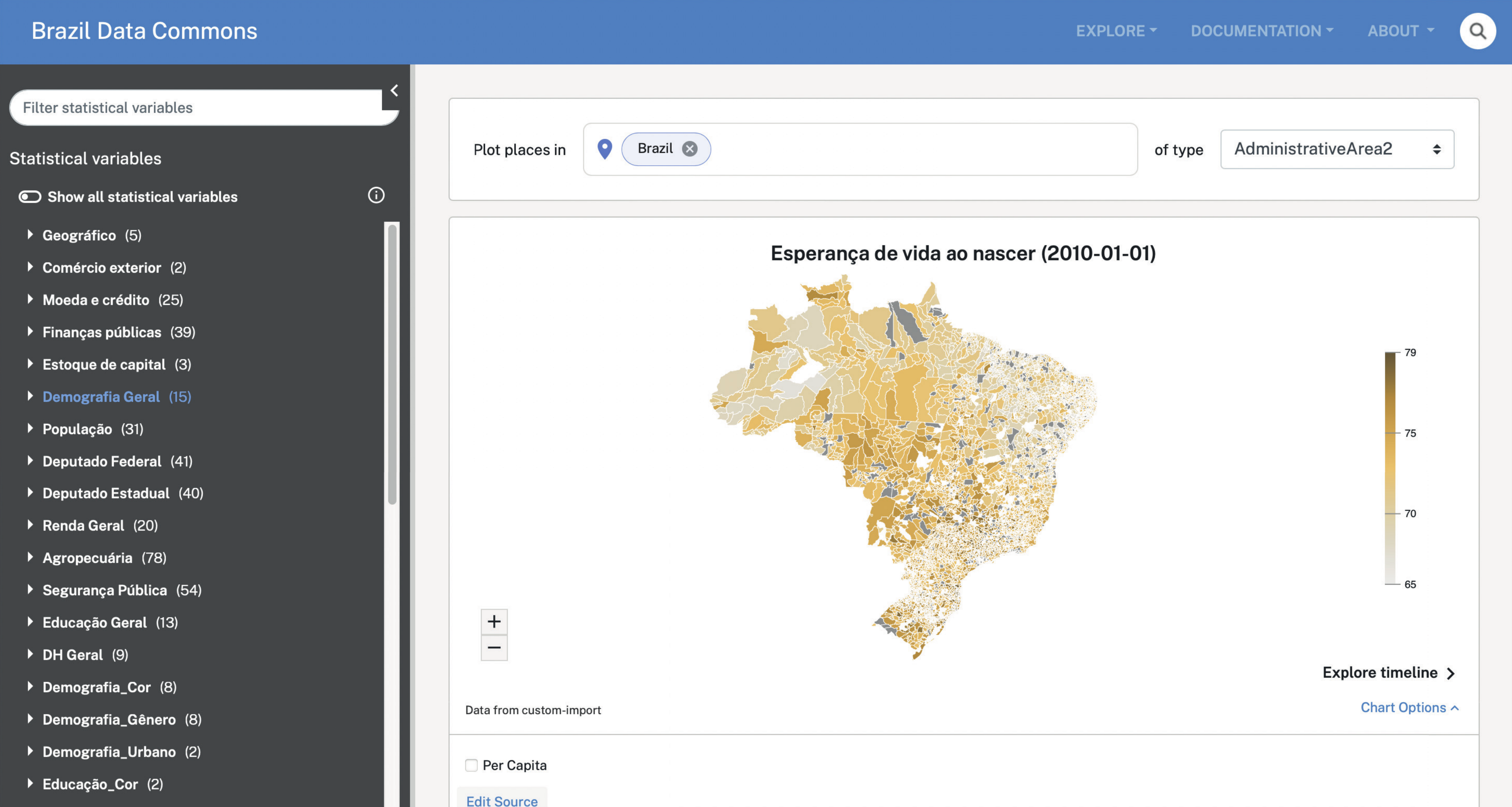}
    \caption{Screenshot of Brazil Data Commons: Map Life Expectancy at Birth for cities in Brazil.}\label{fig:map}
\end{figure}

Figures \ref{fig:timeline}, \ref{fig:scatter}, and \ref{fig:map} display examples of the visualization tools available on the platform. Figure \ref{fig:timeline} illustrates a scenario where the user selects the `Timelines Explorer'' option. After being directed to a dedicated dashboard, users enter the name of a geographic entity (e.g., a country, municipality, or administrative region) to access the datasets available for that location. They can then choose the aggregation level from the drop-down menu on the left panel and select a statistical variable on the right. The resulting visualization, as shown in the example, displays the evolution of Brazilian exports over time. 

Figure \ref{fig:scatter} presents a scenario using the ``Scatter Plot Explorer''. Unlike the timeline visualization, here two statistical variables must be selected—one for the x-axis and another for the y-axis. In this case, the example compares `Total Fertility Rate' and `Employed Women' across Brazilian states, with the state of Alagoas highlighted for reference.

Finally, Figure \ref{fig:map} illustrates the ``Map Explorer''. Similar to the timeline tool, users select a single statistical variable, which is then displayed spatially across the chosen territorial units. In this example, the map represents Life Expectancy at Birth across Brazilian municipalities. Altogether, the three visualization tools support a wide range of customization options, including swapping variables between axes, exploring territorial dimensions, and adding labels to data points.

\begin{figure}[h]
    \centering
    \includegraphics[width=\linewidth]{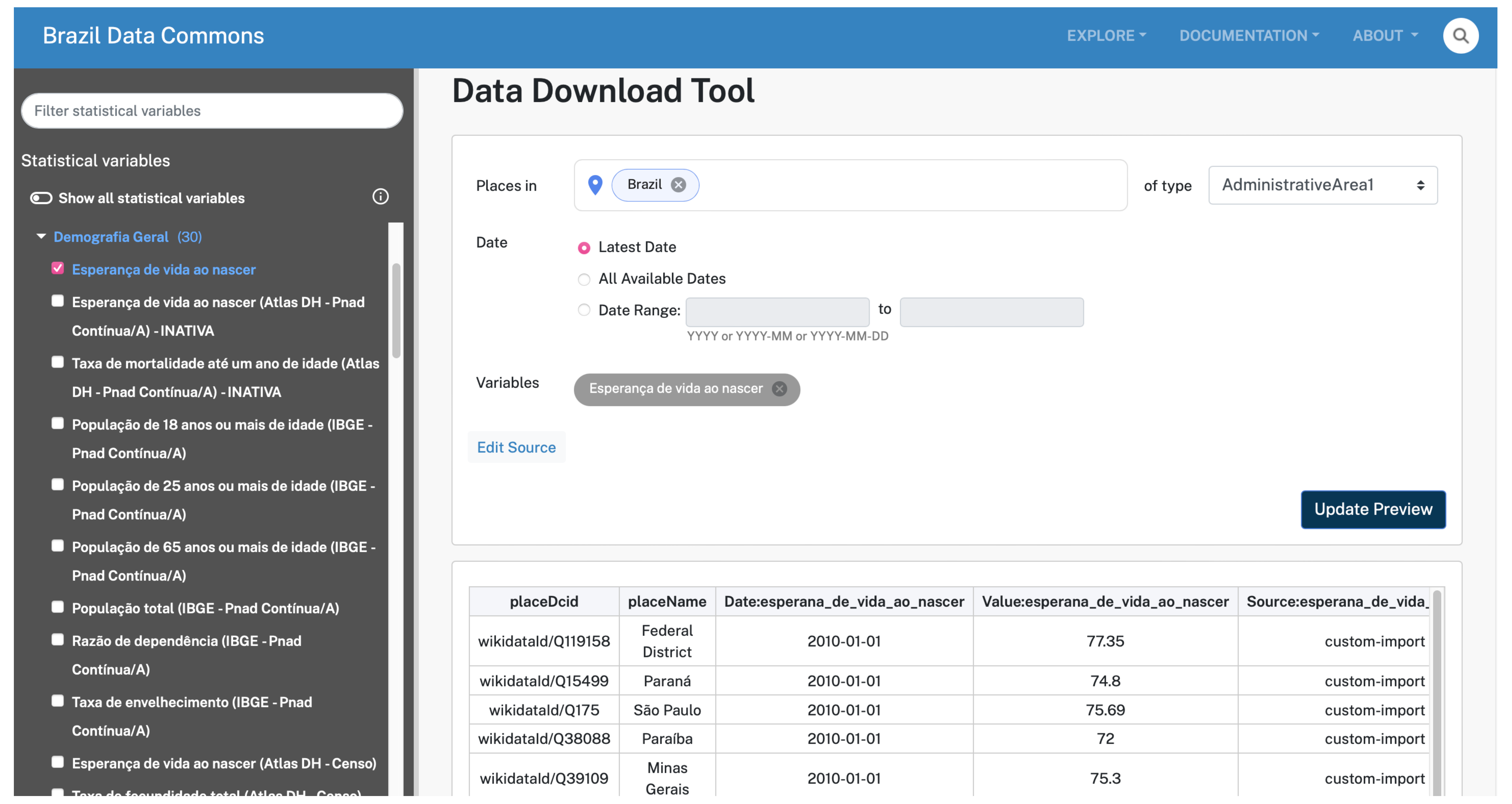}
    \caption{Screenshot of Brazil Data Commons: Download Tool.}\label{fig:downloadtool}
\end{figure}

Another important feature provided by Brazil Data Commons is the ``Data Download Tool'' (Figure \ref{fig:downloadtool}). Through this interface, users can access the full range of datasets available on the platform. As with the visualization tools, the workflow begins by selecting a geographic entity to display the available statistical variables. Users may further customize properties such as the time range and preview the dataset before downloading it. The data can then be exported in \texttt{.csv} format with a single click, ensuring transparency and interoperability for diverse analytical needs.

\subsection{Use Cases}

This subsection presents some potential case studies that can be conducted with the data available in Brazil DC, which can be related to the data available in the global Data Commons.

\subsubsection{Basic Descriptive Analyses}

A fundamental application of the Brazil Data Commons is conducting basic descriptive analyses, which serve as a first step toward understanding key socioeconomic dynamics. One illustrative example is the analysis of monthly labor income trends in Brazil. By exploring the evolution of wages over time, researchers and policymakers can detect inflection points related to economic cycles, government interventions, or external shocks. For instance, the homepage figure displays the significant shifts in income levels during the COVID-19 pandemic, revealing not only the overall income drop at the onset of the crisis but also the subsequent—and uneven—recovery across different demographic groups and regions. Analysts can disaggregate the data by sex, race, region, or age group to further uncover patterns of inequality and resilience in the labor market. These basic insights can support targeted public policies, such as income support programs or training initiatives.

\subsubsection{Spatial Analysis at the Local Level}

Unlike many international datasets that aggregate information at the national or state level, Brazil Data Commons enables fine-grained spatial analysis with data available for over 5,000 Brazilian municipalities. A prime example is the mapping of life expectancy at birth — a synthetic indicator that reflects both health outcomes and socioeconomic development. By examining the spatial distribution of life expectancy, users can identify persistent regional disparities, including stark contrasts between urban centers and underserved rural areas or regions historically marked by structural inequality. This allows for targeted interventions in public health, infrastructure, and education. In addition, local governments and civil society organizations can use this information to prioritize investments and monitor the effects of public policies over time. Through the Brazil Data Commons interface, such indicators can be easily visualized in interactive maps, facilitating interpretation and comparison.

\subsubsection{Data Visualization and Comparison}

One of the core features of Brazil Data Commons is its support for intuitive data visualizations that make complex patterns accessible to non-specialists. For instance, the system enables users to compare the evolution of education indicators—such as literacy rates or highest level of education attained—across countries and time periods. A user may visualize the percentage of adults completing secondary education in Brazil from 2000 to 2022 and compare it with trends in Argentina, South Korea, and South Africa. Such visual comparisons help highlight Brazil’s progress and its remaining challenges in relation to global benchmarks. By reducing the technical barrier to working with time series and international datasets, Brazil Data Commons enables broader engagement with development questions among educators, journalists, students, and policy actors.

\subsubsection{International Comparison}

Because Brazil Data Commons adopts standardized ontologies compatible with the global Data Commons infrastructure, Brazilian datasets can be directly compared with those from other countries. This capability opens up new opportunities for cross-national studies and benchmarking. A case in point is the comparison of average years of schooling across countries. By aligning educational attainment data from Brazil with those from the World Bank or Eurostat, researchers can quantitatively assess Brazil’s position within global educational hierarchies. Such analyses reveal, for example, that Brazil continues to trail behind peer economies in Latin America and far behind high-income countries, particularly in rural and low-income regions. These findings can inform international cooperation efforts and justify investments in education as a lever for inclusive development.

Beyond the specific use cases presented above, the impact of Brazil Data Commons lies, at least partially, in its ability to address structural challenges historically associated with public data in Brazil. Traditionally, working with heterogeneous administrative records has required extensive effort and at least a minimum level of technical expertise, limiting broader use. By integrating these sources, Brazil Data Commons facilitates comparative analyses across countries and regions and supports fine-grained studies at the municipal level, thereby expanding opportunities for evidence-based research.

Importantly, greater accessibility and visibility may in turn contribute to improving the quality of administrative records themselves. Broader use increases scrutiny and creates stronger incentives for institutional refinement. However, current limitations must be acknowledged. For example, mortality and fertility records require preliminary quality assessments, as coverage remains uneven—though more recent periods tend to show higher quality, particularly for longitudinal and regional analyses. These challenges are not restricted to vital statistics; other administrative records also present issues that must be carefully considered.
In this sense, Brazil Data Commons contributes to strengthening the transparency and long-term value of Brazil's public data ecosystem for research and policy.

\section{Ethical Considerations and Responsible Data Usage}\label{sec5}

When individuals share personal data with institutions, they expect that such sensitive information will not be freely disclosed to other organizations or individuals. The data curator, i.e., the entity responsible for collecting and aggregating individuals’ data, must comply with applicable legislation throughout the entire process of data collection, processing, and disclosure. Current legislation around the world represents an effort to ensure that privacy protection is guaranteed for all individuals.

In this section, we examine the key aspects of the responsibilities of data curators in handling personal data, as well as the challenges between transparency and privacy protection.

\subsection{Compliance with privacy and transparency laws}\label{sec7:sub1}

In 2014, the United Nations (UN) General Assembly adopted Resolution 68/261, which establishes a set of fundamental principles for official statistics~\cite{UN-2014}. This resolution determines that individual data collected by statistical agencies, whether relating to individuals or legal entities, must be kept strictly confidential and used exclusively for statistical purposes. Indeed, there are legislative and regulatory efforts worldwide to protect citizens' privacy, such as the General Data Protection Regulation (GDPR) in the European Union~\cite{gdpr}, the United States Confidential Information Protection and Statistical Efficiency Act (CIPSEA)~\cite{US02}, and the Australian revision of its Privacy Act~\cite{AU1988}. 

In Brazil, the data protection law, namely LGPD~\cite{BrazilLGPD2018}, enacted in 2018, establishes sanctions for public authorities in cases where citizens’ privacy is not safeguarded in statistical data disclosures. 
On the other hand, the LAI~\cite{BrazilLAI2011}, in its Article 6, establishes that public bodies and entities shall ensure transparent management of information, providing broad access to it and its dissemination. Therefore, public agencies that collect, maintain, and disclose statistical data face the challenge of reconciling transparency requirements with the privacy protection requirements imposed by current Brazilian legislation.

In the next section we present the main strategies proposed in the literature to diminish privacy risks.

\subsection{Strategies to Mitigate Privacy Risks}

Statistical Disclosure Control (SDC) is field of research that deals with the challenging known trade-off between privacy and utility~\cite{dinur2003revealing, dwork2011firm, fung2010introduction, hundepool2012statistical}: greater privacy assurance for data subjects is generally accompanied by lower data utility for analysts, and vice versa.

Traditionally, institutions have released public data using pseudonymization techniques, which involve removing direct identifiers of individuals (e.g., name, Social Security Number) and claiming that privacy is preserved because the individuals cannot be directly identified. However, studies such as \cite{sweeney2000simple} have shown that knowing only a few attributes about an individual (e.g., sex, age, ZIP code) is often sufficient to re-identify them in a dataset with high probability.

This issue motivated researchers to develop so-called \textit{anonymization} techniques. These methods modify the original dataset through operations such as

\begin{itemize}
    \item \textit{generalization} – reducing the granularity of data, e.g., replacing city values with state values,

    \item \textit{suppression} – removing values considered sensitive, and

    \item \textit{swapping} – exchanging attribute values between individuals.
\end{itemize}

Well-known representatives of anonymization methods include $k$-anonymity and its derivatives ($\ell$-diversity, $t$-closeness, and others). While these methods improve data privacy, they remain vulnerable to compositional attacks, cases where side information from multiple sources is combined by a malicious adversary (a person, or entity) to infer sensitive information about individuals.

To work around the vulnerability present in traditional methods, differential privacy (DP) was proposed in 2006 by Cynthia Dwork~\cite{dwork2006calibrating} and it is still considered the state-of-art in protecting data privacy. The idea of DP is to add calibrated noise to data so the participation of any individual in the dataset does not change by much the released data. So is the problem solved? The answer is no and the motive is that noise addition to data can damage considerably the \textit{utility} of the data, i.e., the purpose that this data was originally collected. The research area in DP and how to balance the trade-off between privacy and utility is very active nowadays with works presenting refinements of DP's original idea.

To address the vulnerabilities of traditional methods, differential privacy (DP) was proposed in 2006 by Cynthia Dwork~\cite{dwork2006calibrating} and is still considered the state-of-art in data privacy protection. The core idea of DP is to add carefully calibrated noise to the data so that the inclusion or exclusion of any individual in the dataset has minimal impact on the released results.

Does this mean the problem is solved? Unfortunately, no. The reason is that adding noise can significantly reduce the utility of the data, i.e., the extent to which the data can still serve its original purpose. Research on DP remains highly active, with ongoing work seeking to refine the original concept and to better balance the trade-off between privacy and utility.

\subsection{Dealing with Microdata}\label{subsec2}
In addition to the inherent trade-off between privacy and utility, an additional challenge arises when the dataset of interest consists of microdata, i.e., individual-level records. The higher the level of granularity, the more difficult it becomes to simultaneously preserve privacy and maintain acceptable utility. For instance, consider a survey aimed at estimating the average height within a population. In this case, the data curator collects individual height measurements, computes the mean, and releases this aggregate value. As discussed previously, one common DP approach is to perturb the reported mean by adding a carefully calibrated random noise. However, if the objective is instead to release the full microdata, that is, the height of each individual, then DP requires injecting noise into every single data point (in the literature it is known by \textit{Local Differential Privacy - LDP}). This per-record perturbation leads to a substantially larger distortion in statistical measurements obtained from this dataset, thereby exacerbating the trade-off between privacy protection and statistical accuracy.

In Brazil, there is a longstanding tradition of transparency regarding data produced by public institutions. As noted in Section~\ref{sec7:sub1}, these institutions are legally bound by LAI \cite{BrazilLAI2011}, which mandates a high degree of openness in the dissemination of collected and processed data. An example of such transparency is the annual release of the Brazilian Educational Censuses by the INEP \cite{inep2025}. The censuses datasets are published in microdata format and contain information on millions of individuals, including students in basic and higher education, participants in the \textit{Exame Nacional do Ensino Médio} (ENEM), a nationwide examination widely used for university admissions), among others. In 2022, Inep, in collaboration with the \textit{Universidade Federal de Minas Gerais} (UFMG), published a study\footnote{The study and accompanying technical note are available online, in Portuguese, at \url{https://www.gov.br/inep/pt-br/centrais-de-conteudo/noticias/institucional/nota-de-esclarecimento-divulgacao-dos-microdados}.} analyzing the privacy risks associated with these microdata releases. The study highlights the challenges of safeguarding privacy in large-scale datasets and emphasizes that, compared to countries such as the United States, Australia, and the Netherlands, Brazil disseminates educational data with a significantly higher degree of granularity and detail, thereby increasing the potential vulnerability to privacy breaches.

DATASUS provides anonymized microdata encompassing vital statistics (births and deaths), hospital admissions, morbidity, and health expenditures. These datasets constitute a fundamental resource for evidence-based public health analysis, enabling assessments of the population’s health status, the evaluation of healthcare system performance, and the formulation of targeted policies and interventions. The systematic measurement of population health has been a longstanding tradition in Brazilian public health, initially established through the continuous recording of mortality and survival data—collectively referred to as Vital Statistics (Mortality and Live Births). The level of granularity varies according to the type of record. For example, death certificates include variables such as sex, age, cause of death according to the International Classification of Diseases (ICD), educational attainment, race or color, occupation, date of death, and municipality of occurrence. Birth records, in turn, contain detailed information on both the newborn and the mother. Quantitatively, these microdata represent the total number of events observed annually, amounting to millions of individual-level entries collected nationwide.

The IBGE \cite{ibge2025} also disseminates extensive collections of microdata, primarily derived from its continuous sample surveys—such as the Continuous National Household Sample Survey (PNAD Contínua)—and from the decennial Population Census. The PNAD Contínua comprises a probabilistic sample of approximately 211,000 households distributed across roughly 3,000 municipalities and adopts a rotating panel design, whereby each household is visited five times within a twelve-month period. The Population Census, conducted every ten years, administers two questionnaires: a short form applied universally and a long form administered to a random and independent subsample within each census tract. Both instruments capture detailed socioeconomic, demographic, and housing characteristics, producing microdata collections that encompass tens of millions of records and serve as an essential empirical foundation for socioeconomic research and public policy design in Brazil.

\section{Conclusion}\label{sec6}

This paper introduces the Brazil Data Commons platform as an effort towards democratizing access to public data in Brazil. By offering an accessible interface and low entry barriers for users of varying technical backgrounds, Brazil Data Commons empowers a broad range of actors—from specialists to the general public—to explore a comprehensive and structured dataset landscape. The platform significantly reduces the time investment traditionally required to access and prepare Brazilian administrative data. Its central contribution lies in enabling comparative analyses across time periods, regions, and even countries, while also supporting granular investigations at subnational levels, particularly within municipalities. We hope that tasks that usually demand months of work to organize datasets from disparate administrative units or to assemble consistent time series for small-area analyses can now be accomplished efficiently through the platform’s integrated and standardized interface, enabling researchers to focus their time on analysis and interpretation rather than data acquisition and preprocessing. 

Beyond the platform itself, a major contribution lies in its promotion of well-defined semantic standards that enable the integration of heterogeneous datasets. A persistent challenge within the Brazilian administrative data landscape has been the lack of interoperability among sources: registries frequently adopt inconsistent variable definitions and identifiers, rendering cross-dataset analyses labor-intensive and accessible only to users with some programming expertise. Brazil Data Commons addresses this structural limitation by functioning as a public good that mitigates these barriers. Through the establishment and implementation of shared standards for data representation, the platform advances not only technical interoperability but also transparent, open, and privacy-conscious data governance practices—principles that are increasingly essential in an era of heightened digital scrutiny and growing concerns about data misuse.

As future work we intend to expand Brazil Data Commons into a more collaborative ecosystem by inviting contributions from external users, researchers, and institutions. Through a participatory model, the platform seeks to broaden the diversity and depth of the available datasets while ensuring its ongoing relevance. An additional, albeit indirect, outcome is the potential improvement in the quality of administrative data for research and public policy analysis. Currently, substantial disparities exist currently between data sources and, by facilitating access and analysis, Brazil Data Commons can help improve the reliability and consistency of such information in the future \cite{tejedo2025}. 

In addition, community engagement fosters shared responsibility in the preservation and maintenance of public data, improving metadata accuracy, identifying data gaps, and encouraging interdisciplinary applications. Within this shared responsibility, the Brazil Data Commons pipeline operates under the principle of autonomy and accountability. Although its architecture grants national teams the flexibility to manage and prepare datasets, it also incorporates a dedicated privacy verification step to protect sensitive and personal information. To the best of our efforts, this process aligns with the Brazilian data protection law~\cite{BrazilLGPD2018} and aims to release only data considered safe for public use. However, privacy risk assessment is inherently contextual and probabilistic: passing such verification does not imply that a dataset is fully immune to re-identification nor to attribute inference. Ideally, this process should involve an interdisciplinary team combining technical, ethical and legal expertise to ensure a balanced approach between data accessibility and individual rights protection. This participatory architecture aligns with the broader vision of data commons as dynamic, evolving infrastructures shaped not only by technology but by the communities that use and contribute to them. 

Finally, we hope the Brazil’s experience may provide a valuable reference for other countries in the Global South, exemplifying how open data infrastructures can serve not only as technical platforms but also as instruments of democratic governance and collective intelligence—empowering societies to transform raw information into knowledge that informs policy, strengthens accountability, and advances inclusive development. In this context, investments in accessible and well-curated public data infrastructures are critical to advancing development agendas, and a growing base of informed data users can, in turn, contribute to improving the quality and accountability of administrative data systems.

\section{Declarations}\label{sec7}

\subsection{Availability of data and materials}\label{subsec3}

All code developed for the Brazil Data Commons system, including both the web architecture and the data processing pipeline, is openly available at: \url{https://github.com/iscris/data-commons-brasil}.

The datasets integrated into the platform can be accessed through their original sources:
\begin{itemize}
    \item Ipeadata: \url{http://www.ipeadata.gov.br/Default.aspx}
    \item IBGE API: \url{https://servicodados.ibge.gov.br/api/docs/}
    \item OpenDataSUS (Ministry of Health): \url{https://opendatasus.saude.gov.br/organization/ministerio-da-saude}
\end{itemize}

\subsection{Competing interests}\label{subsec4}

The authors declare that they have no competing interests.

\subsection{Funding}\label{subsec5}

This work was partially funded by Google.

\subsection{Authors' contributions}\label{subsec6}

All authors contributed equally to the conception of this study. IR, JR, BQ, and FB were involved in designing the platform. IR planned and implemented the platform, while IR and JS handled data extraction and formatting. IR, RG, and MA performed the microdata verification. IR served as the lead author and primary contributor. All authors reviewed and approved the final version of the manuscript.

\subsection{Acknowledgements}\label{subsec7}
Not applicable.


\bibliography{sn-bibliography}

@inproceedings{gonze2025riscos,
  title={Riscos de Privacidade em Dados de Sa{\'u}de: Investigando Infer{\^e}ncia de Atributos Sens{\'\i}veis de Cidad{\~a}os no DATASUS},
  author={Gonze, Ramon G and Lemes, Igor W and Almeida, Jussara M and Gon{\c{c}}alves, Marcos A and Alvim, M{\'a}rio S},
  booktitle={Simp{\'o}sio Brasileiro de Seguran{\c{c}}a da Informa{\c{c}}{\~a}o e de Sistemas Computacionais (SBSeg)},
  pages={790--806},
  year={2025},
  organization={SBC}
}

@inproceedings{dwork2006calibrating,
  title={Calibrating noise to sensitivity in private data analysis},
  author={Dwork, Cynthia and McSherry, Frank and Nissim, Kobbi and Smith, Adam},
  booktitle={Theory of cryptography conference},
  pages={265--284},
  year={2006},
  organization={Springer}
}

@article{sweeney2000simple,
  title={Simple demographics often identify people uniquely},
  author={Sweeney, Latanya},
  journal={Health (San Francisco)},
  volume={671},
  number={2000},
  pages={1--34},
  year={2000}
}

@book{hundepool2012statistical,
  title={Statistical disclosure control},
  author={Hundepool, Anco and Domingo-Ferrer, Josep and Franconi, Luisa and Giessing, Sarah and Nordholt, Eric Schulte and Spicer, Keith and De Wolf, Peter-Paul},
  year={2012},
  publisher={John Wiley \& Sons},
  address={Statistics Netherlands, The Netherlands}
}

@article{dwork2011firm,
	title={A firm foundation for private data analysis},
	author={Dwork, Cynthia},
	journal={Communications of the ACM},
	volume={54},
	number={1},
	pages={86--95},
	year={2011},
	publisher={ACM New York, NY, USA}
}

@book{fung2010introduction,
	title={Introduction to privacy-preserving data publishing: Concepts and techniques},
	author={Fung, Benjamin CM and Wang, Ke and Fu, Ada Wai-Chee and Philip, S Yu},
	year={2010},
	publisher={Chapman and Hall/CRC},
    address={Minneapolis, The United States of America}
}

@inproceedings{dinur2003revealing,
	title={Revealing information while preserving privacy},
	author={Dinur, Irit and Nissim, Kobbi},
	booktitle={Proceedings of the twenty-second ACM SIGMOD-SIGACT-SIGART symposium on Principles of database systems},
	pages={202--210},
	year={2003}
}

@misc{BrazilLAI2011,
  author       = {{Brazil}},
  title        = {Lei de Acesso à Informação (LAI), Law No. 12.527},
  year         = {2011},
  howpublished = {\url{https://www.planalto.gov.br/ccivil_03/_ato2011-2014/2011/lei/l12527.htm}},
  note         = {Enacted 18 November 2011}
}

@misc{BrazilLGPD2018,
  author       = {Brazil},
  title        = {Lei Geral de Proteção de Dados Pessoais (LGPD), Law No. 13.709},
  year         = {2018},
  howpublished = {\url{https://www.planalto.gov.br/ccivil_03/_ato2015-2018/2018/lei/l13709.htm}},
  note         = {Enacted 14 August 2018, amended by Law No. 13.853 of 8 July 2019}
}

@misc{AU1988,
	author = {Government of Australia},
	howpublished = {\url{https://www.legislation.gov.au/Details/C2015C00598}},
	title = {{Privacy Act 1988}},
	year = {1988}
}

@misc{US02,
	author = {Government of the United States of America},
	howpublished = {\url{https://www.eia.gov/cipsea/cipsea.pdf}},
	title = {Confidential Information Protection and Statistical Efficiency Act (CIPSEA)},
	year = {2002}
}

@misc{gdpr,
    title={Regulation (EU) 2016/679 of the European Parliament and of the Council of 27 April 2016 on the protection of natural persons with regard to the processing of personal data and on the free movement of such data, and repealing Directive 95/46/EC (General Data Protection Regulation)},
    author={EU},
    year={2016},
    note={Available at \url{https://eur-lex.europa.eu/eli/reg/2016/679/oj}}
}

@misc{UN-2014,
	author="{Organização das Nações Unidas}",
	title="Fundamental {P}rinciples of {O}fficial {S}tatistics ({A}/{RES}/68/261 from 29 {J}anuary 2014)",
	url = "https://unstats.un.org/unsd/dnss/gp/fundprinciples.aspx",
	year = "2014",
	note = {Disponível em: \url{https://unstats.un.org/unsd/dnss/gp/fundprinciples.aspx}.}
}

@article{herrera2007improving,
  author        = "Herrera, Y. M. and Kapur, D.",
  title			= "Improving data quality: Actors, incentives, and                    capabilities",
  journal		= "Political Analysis",
  volume		= "15",
  number        = "4",
  pages			= "365--386",
  year			= "2007",
  doi			= "https://doi.org/10.1093/pan/mpm007",
  publisher     = "Cambridge University Press",
}

@book{shikida2021guia,
  author		= "Shikida, C. D. and Monasterio, L. and Nery, P.                     F.",
  title			= "Guia brasileiro de análise de dados: armadilhas \&                  soluções",
  address		= "Brazil",
  publisher		= "Enap",
  year			= "2021"
}

@misc{passos2022,
  author       = {Passos, J.},
  title        = {Falta de integração e distribuição das bases de dados fragiliza sistemas de informação},
  year         = {2022},
  howpublished = {\url{https://www.epsjv.fiocruz.br/noticias/reportagem/falta-de-integracao-e-distribuicao-das-bases-de-dados-fragiliza-sistemas-de}},
  note         = {Accessed July 30 2025}
}

@article{dang2023statistical,
  author        = "Dang, H. H. and Pullinger, J. and Serajuddin, U.                   and Stacy, B.",
  title			= "Statistical performance indicators and index—a                     new tool to measure country statistical capacity",
  journal		= "Scientific Data",
  volume		= "10",
  number        = "146",
  year			= "2023",
  doi			= "https://doi.org/10.1038/s41597-023-01971-0",
  publisher     = "Springer Nature",
}

@misc{press2016,
  author       = {Press, G.},
  title        = {Data Preparation: Most Time-Consuming, Least Enjoyable Data Science Task, Survey Says},
  year         = {2016},
  howpublished = {\url{https://www.forbes.com/sites/gilpress/2016/03/23/data-preparation-most-time-consuming-least-enjoyable-data-science-task-survey-says/?sh=41cfc7606f63}},
  note         = {Accessed July 30 2025}
}

@article{martins2013,
  author        = "Martins, S. C. and Mauritti, R. and da Costa, A. F.",
  title			= "Acesso a bases de microdados: aplica{\c{c}}                        {\~o}es e impactos nas pesquisas em ci{\^e}ncias                   sociais",
  journal		= "Media{\c{c}}{\~o}es-Revista de Ci{\^e}ncias                        Sociais",
  volume		= "18",
  number        = "1",
  year			= "2013",
  doi			= "https://doi.org/10.5433/2176-6665.2013v18n1p66",
}

@misc{guha2023,
  author        = "Guha, R. V. and Radhakrishnan, P. and Xu, B. and                   Sun, W. and Au, C. and Tirumali, A. and Amjad, M.                  J. and Piekos, S. and Diaz, N. and Chen, J. and                    others",
  title         = "Data commons",
  year          = "2023",
  note          = "Preprint at \url{https://arxiv.org/abs/2309.13054}"
}

@misc{schemaorg2015,
  title         = "Schema.org",
  year          = "2015",
  note          = "\url{https://schema.org}",
  howpublished  = "Accessed July 30 2025"
}

@article{blicharska2017,
  author        = "Blicharska, M. and Smithers, R. J. and Kuchler, M. and Agrawal, G. K. and Gutiérrez, J. M. and Hassanali, A. and Huq, S. and Koller, S. H. and Marjit, S. and Mshinda, H. M. and others",
  title         = "Steps to overcome the North--South divide in research relevant to climate change policy and practice",
  journal       = "Nature Climate Change",
  volume        = "7",
  number        = "1",
  pages         = "21--27",
  year          = "2017",
  doi           = "https://doi.org/10.1038/nclimate3163",
  publisher     = "Nature Publishing Group UK London"
}

@article{coelho2021,
  author        = "Coelho Neto, G. C. and Chioro, A.",
  title         = "Afinal, quantos Sistemas de Informação em Saúde de base nacional existem no Brasil?",
  journal       = "Cadernos de Saúde Pública",
  volume        = "37",
  pages         = "e00182119",
  year          = "2021",
  doi           = "https://doi.org/10.1590/0102-311X00182119",
  publisher     = "SciELO Public Health",
}

@misc{saliba2023,
  author       = {Saliba, P.},
  title        = {Privacy, public transparency, and climate change},
  year         = {2023},
  howpublished = {\url{https://medium.com/opendatacharter/privacy-public-transparency-and-climate-change-04f671a2683c}},
  note         = {Accessed July 30 2025}
}

@inproceedings{deoliveira2018,
  author        = "de Oliveira, E. F. and Silveira, M. S.",
  title         = "Open Government Data in Brazil: A Systematic                       Review of its Uses and Issues",
  booktitle     = "Proceedings of the 19th Annual International                       Conference on Digital Government Research (dg.o                    '18)",
  series        = "ACM International Conference Proceeding Series",
  pages         = "1--9",
  year          = "2018",
  address       = "Delft, Netherlands",
  publisher     = "Association for Computing Machinery",
  doi           = "https://doi.org/10.1145/3209281.3209339",
}

@inproceedings{correa2014,
  author        = "Corrêa, A. S. and Corrêa, P. L. P. and da                        Silva, F. S. C.",
  title         = "Transparency Portals versus Open Government                      Data: An Assessment of Openness in Brazilian                     Municipalities",
  booktitle     = "Proceedings of the 15th Annual International                     Conference on Digital Government Research",
  pages         = "178--185",
  year          = "2014",
  address       = "Aguascalientes, Mexico",
  publisher     = "Association for Computing Machinery",
}

@article{breitman2012,
  author        = "Breitman, K. and Salas, P. and Casanova, M. A. and Saraiva, D. and Gama, V. and Viterbo, J. and Magalhães, R. P. and Franzosi, E. and Chaves, M.",
  title         = "Open Government Data in Brazil",
  journal       = "IEEE Intelligent Systems",
  volume        = "27",
  number        = "3",
  pages         = "45--49",
  year          = "2012",
  publisher     = "IEEE Computer Society",
  doi           = "https://doi.org/10.1109/MIS.2012.25",
}

@article{tejedo2025,
  author        = "Tejedo-Romero, F. and Ferraz, J. F. E. A. and                      Ribeiro, M. J. G.",
  title         = "The usability of Brazilian government open data                    portals: ensuring data quality",
  journal       = "Humanities and Social Sciences Communications",
  volume        = "12",
  number        = "1",
  pages         = "1--13",
  year          = "2025",
  publisher     = "Palgrave",
  doi           = "https://doi.org/10.1057/s41599-025-04404-y",
}

@techreport{freitas2023,
  author       = "Freitas, E. E. and Romero, J. P. and Britto, G. and Stein, A. Q. and Torres, R.",
  title        = "Dataviva: espaço de atividades e indicadores regionais de complexidade econômica",
  institution  = "Cedeplar, Universidade Federal de Minas Gerais",
  year         = "2023",
}

@article{guha2016,
  author        = "Guha, R. V. and Brickley, D. and Macbeth, S.",
  title         = "Schema.org: evolution of structured data on the web",
  journal       = "Communications of the ACM",
  volume        = "59",
  number        = "2",
  pages         = "44--51",
  year          = "2016",
  publisher     = "ACM",
  doi           = "https://doi.org/10.1145/2844544",
}

@misc{brickley1998,
  author       = {Brickley, D. and Guha, R. V.},
  title        = {Resource Description Framework Schema},
  institution  = {World Wide Web Consortium},
  year         = {1998},
  howpublished = {\url{https://www.w3.org/TR/WD-rdf-schema/}},
  note         = {Accessed on August 19, 2025}
}

@misc{undp2025,
  author       = "United Nations Development Programme (UNDP)",
  title        = "Human Development Index (HDI)",
  institution  = "Human Development Report Office (HDRO)",
  year         = "2025",
  howpublished = "\url{https://hdr.undp.org/data-center/human-development-index}",
  note         = "Accessed August 19 2025"
}

@misc{guha2015,
  author        = "Guha, R. V. and Gupta, V.",
  title         = "Communicating semantics: Reference by description",
  year          = "2015",
  note          = "Preprint at \url{https://arXiv:1511.06341}",
  doi           = "https://doi.org/10.48550/arXiv.1511.06341",
}

@article{bertot2010,
  author        = "Bertot, J. C. and Jaeger, P. T. and Grimes, J. M.",
  title         = "Using ICTs to create a culture of transparency: E-government and social media as openness and anti-corruption tools for societies",
  journal       = "Government Information Quarterly",
  volume        = "27",
  number        = "3",
  pages         = "264--271",
  year          = "2010",
  publisher     = "Elsevier",
  doi           = "https://doi.org/10.1016/j.giq.2010.03.001",
}

@article{lourenco2015,
  author        = "Lourenço, R. P.",
  title         = "An analysis of open government portals: A perspective of transparency for accountability",
  journal       = "Government Information Quarterly",
  volume        = "32",
  number        = "3",
  pages         = "323--332",
  year          = "2015",
  publisher     = "Elsevier",
  doi           = "https://doi.org/10.1016/j.giq.2015.05.006",
}

@misc{datagov2025,
  author       = {{United States Government}},
  title        = "Data.gov",
  institution  = "U.S. Government",
  howpublished = "\url{https://www.data.gov/}",
  note         = "Accessed September 1 2025",
}

@misc{datagovuk2025,
  author       = {{United Kingdom Government}},
  title        = "Data.gov.uk",
  institution  = "UK Government",
  howpublished = "\url{https://www.data.gov.uk/}",
  note         = "Accessed September 1 2025",
}

@misc{dadosgovbr2025,
  author       = {{Governo Federal do Brasil}},
  title        = "Dados.gov.br",
  institution  = "Governo Federal",
  howpublished = "\url{https://dados.gov.br/}",
  note         = "Accessed September 1 2025",
}

@misc{basedosdados2025,
  title        = "Base dos Dados",
  howpublished = "\url{https://basedosdados.org}",
  note         = "Accessed September 3 2025",
}

@misc{pdd2025,
  author = "Cidacs",
  title = "Plataforma de Dados Desidentificados",
  howpublished = "\url{https://pdd.cidacs.org/}",
  note = "Accessed September 3 2025",
}

@misc{worldbank2025,
  author       = {{World Bank Group}},
  title        = "World Bank Open Data",
  howpublished = "\url{https://data.worldbank.org}",
  note         = "Accessed September 3 2025",
}

@misc{oecd2025,
  author       = {{Organisation for Economic Co-operation and Development}},
  title        = "OECD",
  howpublished = "\url{https://www.oecd.org/en.html}",
  note         = "Accessed September 3 2025",
}

@misc{eurostat2025,
  title        = {{Eurostat: Statistical database of the European Union}},
  howpublished = "\url{https://ec.europa.eu/eurostat}",
  note         = "Accessed September 3 2025",
}

@misc{ipea2019,
  author       = {{Instituto de Pesquisa Econômica Aplicada (IPEA)}},
  title        = "Ipeadata",
  howpublished = "\url{http://www.ipeadata.gov.br/Default.aspx}",
  note         = "Accessed September 3 2025",
}

@misc{ibge2025,
  title        = "Instituto Brasileiro de Geografia e Estatística (IBGE)",
  howpublished = "\url{https://www.ibge.gov.br}",
  note         = "Accessed September 3 2025",
}

@misc{datasus2025,
  author       = {{Departamento de Informática do Sistema Único de Saúde}},
  title        = "DATASUS: Banco de dados do Sistema Único de Saúde",
  howpublished = "\url{https://datasus.saude.gov.br}",
  note         = "Accessed September 3 2025",
}

@misc{sinan2025,
  title        = "Sistema de Informação de Agravos de Notificação (SINAN)",
  howpublished = "\url{https://portalsinan.saude.gov.br}",
  note         = "Accessed October 30 2025",
}

@misc{sipni2025,
  author       = {{Ministério da Saúde}},
  title        = "SI-PNI: Programa Nacional de Imunizações",
  howpublished = "\url{https://si-pni.saude.gov.br/#/login}",
  note         = "Accessed October 30 2025",
}

@misc{sisagua2025,
  author       = {{Ministério da Saúde}},
  title        = "Sistema de Informação de Vigilância da Qualidade da Água para Consumo Humano (SISAGUA)",
  howpublished = "\url{http://sisagua.saude.gov.br/sisagua/paginaExterna.jsf}",
  note         = "Accessed October 30 2025",
}

@misc{inep2025,
  title        = "Instituto Nacional de Estudos e Pesquisas Educacionais Anísio Teixeira (INEP)",
  howpublished = "\url{https://www.gov.br/inep/pt-br}",
  note         = "Accessed October 27 2025",
}

@misc{google2025,
  title        = "Google",
  howpublished = "\url{https://www.google.com/}",
  note         = "Accessed September 3 2025",
}

@misc{microsoft2025,
  title        = "Microsoft",
  howpublished = "\url{https://www.microsoft.com/pt-br/}",
  note         = "Accessed September 3 2025",
}

@misc{pinterest2025,
  title        = "Pinterest",
  howpublished = "\url{https://br.pinterest.com}",
  note         = "Accessed September 3 2025",
}

@misc{yandex2025,
  title        = "Yandex",
  howpublished = "\url{https://yandex.com}",
  note         = "Accessed September 3 2025",
}

@techreport{alvim2020, 
  author = "Alvim, M. S. and van de Graaf, J. and Gonze, R. G. and Nunes, G. H.",
  title = "TED 8750 - PRICE: Privacidade nos Censos Educacionais",
  institution = "Universidade Federal de Minas Gerais",
  year = "2020", }

@misc{rnds2025,
  author       = "Ministério da Saúde",
  title        = "Rede Nacional de Dados em Saúde (RNDS)",
  year         = "2020",
  howpublished = "\url{https://www.gov.br/saude/pt-br/composicao/seidigi/rnds}",
  note         = "Accessed on October 25, 2025"
}

@article{silva2025open,
  title={Open Data Observatory in Brazil: An Academic Initiative},
  author={Silva, P. N.},
  journal={Journal of Librarianship and Scholarly Communication},
  volume={12},
  number={2},
  year={2025},
  publisher={Iowa State University Digital Press},
  doi = {https://doi.org/10.31274/jlsc.18312}
}

@inproceedings{toledo2023brstats,
  title={BrStats: a socioeconomic statistics dataset of the Brazilian cities},
  author={Toledo, JM and Moura, Thiago JM and Timoteo, RDA},
  booktitle={Dataset Showcase Workshop (DSW)},
  pages={67--78},
  year={2023},
  organization={SBC}
}

@article{oliveira2024open,
  title={Open government data representation and retrieval: a literatura review},
  author={Oliveira, Danielle Teixeira de and Silva, Patr{\'\i}cia Nascimento},
  journal={RDBCI: Revista Digital de Biblioteconomia e Ci{\^e}ncia da Informa{\c{c}}{\~a}o},
  volume={22},
  pages={e024029},
  year={2024},
  publisher={SciELO Brasil}
}


\end{document}